\PassOptionsToPackage{unicode}{hyperref}
\PassOptionsToPackage{hyphens}{url}
\documentclass[
]{article}
\usepackage{amsmath,amssymb}
\usepackage{iftex}
\ifPDFTeX
  \usepackage[T1]{fontenc}
  \usepackage[utf8]{inputenc}
  \usepackage{textcomp} 
\else 
  \usepackage{unicode-math} 
  \defaultfontfeatures{Scale=MatchLowercase}
  \defaultfontfeatures[\rmfamily]{Ligatures=TeX,Scale=1}
\fi
\usepackage{lmodern}
\ifPDFTeX\else
\fi
\IfFileExists{upquote.sty}{\usepackage{upquote}}{}
\IfFileExists{microtype.sty}{
  \usepackage[]{microtype}
  \UseMicrotypeSet[protrusion]{basicmath} 
}{}
\makeatletter
\@ifundefined{KOMAClassName}{
  \IfFileExists{parskip.sty}{%
    \usepackage{parskip}
  }{
    \setlength{\parindent}{0pt}
    \setlength{\parskip}{6pt plus 2pt minus 1pt}}
}{
  \KOMAoptions{parskip=half}}
\makeatother
\usepackage{xcolor}
\usepackage{longtable,booktabs,array}
\usepackage{multirow}
\usepackage{calc} 
\usepackage{etoolbox}
\makeatletter
\patchcmd\longtable{\par}{\if@noskipsec\mbox{}\fi\par}{}{}
\makeatother
\IfFileExists{footnotehyper.sty}{\usepackage{footnotehyper}}{\usepackage{footnote}}
\makesavenoteenv{longtable}
\usepackage{graphicx}
\makeatletter
\def\maxwidth{\ifdim\Gin@nat@width>\linewidth\linewidth\else\Gin@nat@width\fi}
\def\maxheight{\ifdim\Gin@nat@height>\textheight\textheight\else\Gin@nat@height\fi}
\makeatother
\setkeys{Gin}{width=\maxwidth,height=\maxheight,keepaspectratio}
\makeatletter
\def\fps@figure{htbp}
\makeatother
\ifLuaTeX
  \usepackage{selnolig}  
\fi
\usepackage{bookmark}
\IfFileExists{xurl.sty}{\usepackage{xurl}}{} 
\hypersetup{
  hidelinks,
  pdfcreator={LaTeX via pandoc}}

\author{}
\date{}
\usepackage{caption}
\usepackage{hyperref}

\begin{document}

\title{\Large{\textbf{Designing Rewards for Rewarding Designs:}}\\ \Large{Demonstrating the Impact of Rewards\\on the Creative Design Process}}
\maketitle
\vspace{-1.9cm}
Surabhi S Nath$^{1}$, Vindula Jayawardana$^{2}$, Monica Van$^{2}$, Matt Klenk$^{2}$, and Shabnam Hakimi$^{2}$

\textsuperscript{1} Max Planck Institute for Biological Cybernetics, Tübingen, Germany\\
\textsuperscript{2} Toyota Research Institute, California, USA\\
\vspace{-0.4cm}
\begin{quote}
\textbf{Abstract.} The creative design process involves transforming abstract goals into concrete outcomes through a series of decisions made under constraints. While such processes are commonly shaped by feedback like rewards, their impact on design decision making remains unclear. To better understand the role of rewards in the design process, we modeled a 3D parametric, goal-based chair design task as a Markov Decision Process. We tracked participants' decisions as they iteratively developed designs for an abstract design goal, and presented either a goal-aligned or goal-agnostic reward at every step. We tested the effect of these rewards on task behaviour and self-reported experience. With rewards, participants more thoroughly explored the design space, and maximised goal-aligned over goal-agnostic rewards while preserving diversity across designs. The nature of the goal also mattered, influencing participants' perception of the reward's usefulness. Building on these insights, we propose guidelines for designing effective feedback for design decision making.
\end{quote}


\vspace{-0.35cm}
\section{Introduction}
\vspace{-0.25cm}
Design is a goal-directed sequence of decisions through which designers generate and refine ideas while balancing constraints \cite{simon1973structure,dorst2001creativity}. Feedback is one such constraint: it is integral to the design process and shapes design behaviour in many ways \cite{schon1986reflective}. Design reviews, for example, are central to many design contexts, serving as pivotal checkpoints that reshape the design process and its outcomes \cite{cooper2010stagegate,valeri2003phase}. Feedback further aids designers in refining concepts, mitigating biases, and supporting exploration \cite{fischer1993embedding}.

Consistent with its vital role, feedback has been a key feature of design support tools. For example, a variety of digital tools leverage feedback to encourage ideation and exploration \cite{son2022creativesearch,son2022bigexplore} through multiple modalities, including textual and image-based feedback from generative AI agents \cite{lee2024sketchguided,swearngin2020scout,lee2020guicomp,son2024genquery,lin2025inkspire}. At the same time, not all feedback is beneficial. Feedback can result in design fixation and homogeneity, especially when it is AI-generated \cite{wadinambiarachchi2024effects,anderson2024homogenization}. The proliferation and increasing adoption of such tools underscores the importance of understanding how feedback impacts design behaviour and designers' experience. Although research is beginning to address these issues \cite{bernal2015computational,nandy2023machine}, prior work has largely been either conceptual or purely outcome-focused, motivating the need for fine-grained, mechanistic investigation.

To address this gap, we measured design at a level more closely aligned with its psychological substrates, tracking the process as a series of sequential design decisions and their consequent outcomes. Parametric interfaces provide a natural setting for such investigation, as they are commonly used in professional design practice and represent designs as a set of parameters that can be easily and measurably manipulated. Critically, they produce a stream of intermediate designs that can be logged and analysed \cite{shireen2011design,lee2020creative,cristie2021versioning}. This structure supports a formulation of design as a Markov Decision Process (MDP), with \emph{states} corresponding to intermediate designs, \emph{actions} corresponding to parameter edits, and feedback operationalised as \emph{reward} \cite{lahikainen2024creativity}. This computational account presents testable predictions about the relationship between reward and behaviour, enabling the development of precise, process-level rewards. Such reward reinforcement can augment designers' own intrinsic rewards, and impact learning and decision making at the level of individual actions.

However, designing a \emph{good} reward---a reward that helps improve design outcomes while complementing intrinsic rewards and idiosyncratic preferences---is both essential and far from trivial \cite{ryan2000self,ryan1982control,sutton1998reinforcement}. While rewards can act as informative feedback that steers designers towards goal-relevant designs, they can also become a controlling constraint that collapses exploration, motivates ``reward hacking,'' or be rejected as unhelpful \cite{kluger1996feedback}. These tensions are particularly salient in creative design, where success is rarely defined by a single optimum and multiple distinct solutions can satisfy the same goal \cite{dorst2001creativity}. Systematic investigation can reveal how different forms of reward impact design behaviours, such as exploration, outcome diversity, as well as acceptance and perceived usefulness of feedback.

The consequences of such a mechanistic understanding of the role of feedback in design decision making are far-reaching. For one, they enable the design of better rewards that can facilitate design thinking without hampering creative agency and well-being \cite{ryan2000self}. Second, they enable a window into the cognitive mechanisms underlying design: how designers adapt to constraints, decide which signals to trust, and how they shape the design process \cite{simon1973structure}. Third, from a practical perspective, such insights can inform organisations' review processes to ensure that management provides the right feedback to designers resulting in better functioning design teams and ultimately, better products \cite{cooper2010stagegate}. Finally, and more broadly, principled reward design can guide the development of more effective AI-based design tools that can educate and support the next generation of designers \cite{ivcevic2024ai}.

Here, we systematically investigate the role of rewards in the design process. We posit that an MDP is a fundamentally useful framework for modelling design decision making under rewards. Within this framework, we hypothesize that the reward influences both decision making behaviour and subjective experience, and that the degree of influence depends on the quality of feedback and nature of the goal. Specifically, we expect that rewards will be more useful when they are better aligned with the design goal. Towards this end, our work makes the following contributions:

\begin{itemize}
\item
  Models a 3D parametric goal-based design task as an MDP
\item
  Develops goal-aligned and goal-agnostic reward signals for a large design space and delivers reward feedback at every step of the design process
\item
  Presents empirical data from 353 human participants demonstrating the impact of rewards on design behaviour at multiple levels of analyses and self-reported experience, across three abstract semantic goals
\item
  Concludes with a future-looking guide for effective reward design
\end{itemize}
\vspace{-0.2cm}
\section{Method}

\subsection{Parametric Design Task}
Our task was adapted from the 3D parametric, goal-based chair design environment described by \cite{nandy2025semantic} where participants created chairs by controlling a set of parametric features. Through sequential edits, this environment enabled transparent, process-level insights into design, thus capturing the evolution of the design as well as the underlying decision dynamics.

\textbf{Design Features and Interface.} The parametric design task included 15 features related to the full chair and its components, such as body, legs and arms. Four of these features were primarily aesthetic, namely the chair's material and the colours of each component. The remaining 11 features were functional (\emph{i.e.}, related to chair type and component dimensions). Features could be discrete or continuous. Three discrete features were presented as dropdown menus: leg type (with eight different options, including `no leg'), arm type (with three different options, including `no arm'), and material (with nine different options, including `no material'). Remaining features were continuous in the range of {[}0-1{]}, including all functional attributes and colours, which were controlled using sliders. Each colour feature was input in HSV (hue, saturation, value) format, totalling 18 continuous features. The features were presented in three blocks in a horizontal panel at the top of the screen: type features (arm type, leg type), dimension sliders (for body, arms, and legs), and aesthetic features (colour and material). The six possible block orders were fully randomised across participants. For the starting configuration, all features were initialised with neutral default values: no arms, no legs, no material, slider midpoints for continuous functional features, and a neutral grey default for colour. Save and Reset buttons allowed saving the design or resetting to the starting configuration. The interface was programmed in JavaScript, and the 3D chairs were created using Blender.

\textbf{Goals.} We specified one of three semantic goals (\emph{cheerful}, \emph{dependable}, or \emph{unique)} and participants were instructed to ``design chairs that are \textless goal\textgreater.'' These goals were selected from a larger set to reflect semantic and conceptual diversity and also span a range of imagability and creativity (of the three goals, \emph{unique} was most likely to be associated with creative outcomes \cite{nandy2025semantic}). Besides these abstract, extrinsic goals, we also included a purely intrinsic design goal driven by subjective preferences, where participants were instructed to ``design chairs that you like''.

\textbf{Rewards.} Notably, we introduced reward feedback as a score displayed immediately after each step of the design process. We contrasted a learned, \textbf{goal-aligned} reward signal with a participant-specific, \textbf{goal-agnostic} reward signal with identical structure and complexity as the goal-aligned reward signal. We describe the details of learning and administering the rewards in Section 2.3.

\begin{figure}
    \centering
    \includegraphics[width=0.95\linewidth]{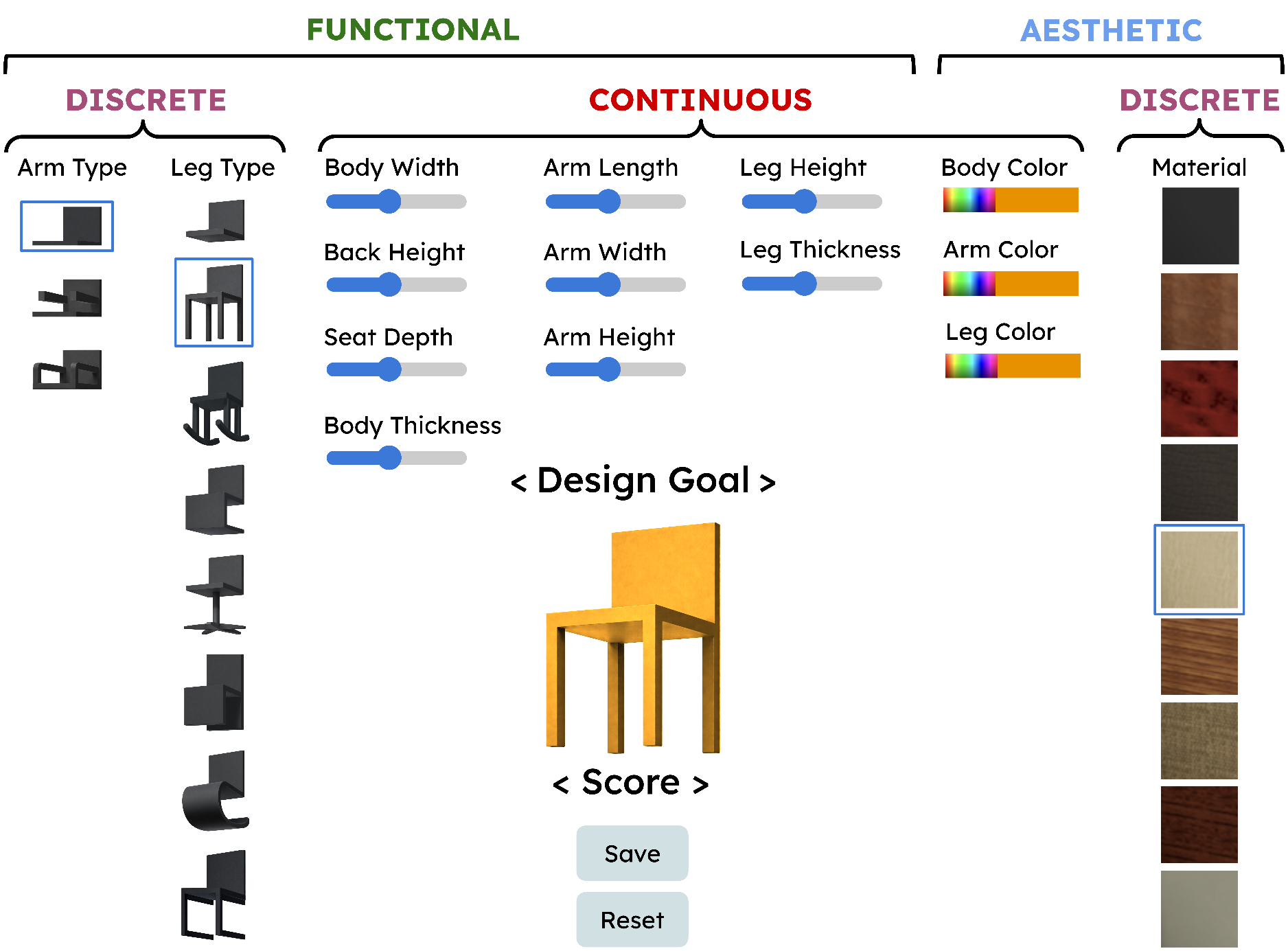}
    \caption{Design environment with all features (within feature categories), sliders, and buttons. Placeholders show the design goal, reward feedback score, and an example chair design.}
    \label{fig:design-environment}
\end{figure}

\autoref{fig:design-environment} depicts the design environment, including all features (with feature categories), buttons, and a sample designed chair. Placeholders for the goal and the reward feedback score are also visualised.

\subsection{Markov Decision Process Formulation}

We formalised the process of designing chairs in this design environment as an MDP \cite{bellman1957mdp}. Let goal \(g\ \epsilon\ G =\) \{\emph{cheerful}, \emph{dependable}, \emph{unique}\}, \(|G| = 3.\) We denoted continuous features as \(c\ \epsilon\ C,\) and discrete features as \(d\ \epsilon\ D.|C| = 18\), \(|D| = 3\) with \(O_{d}\) denoting the set of options for \(d\). For a given participant \(p,\) we defined the chair design MDP as,

\begin{enumerate}
\def\labelenumi{\arabic{enumi}.}
\item
  \(S\) (state space or design space), which denotes the set of states given by the set of chair designs. Each state \(s\ \epsilon\ S\) is represented as a vector\footnote{Formally, the state should include history of previously saved chairs as participants are unlikely to repeat their previous designs. We omit this notation for simplicity.} \(s\  = (x_{1},\ ..x_{|C|},\ z_{1},\ ..z_{|D|})\), where \(x_{c}\epsilon\ \lbrack 0,1\rbrack\), \hspace{0pt}\(z_{d}{\ \epsilon\ O}_{d}\) are the continuous feature values and discrete options of \(s\). \(s_{0}\) denotes the starting configuration.
\item
  \(A\) (action space), which denotes the set of actions. \(a\ \epsilon\ A = \ (c,x_{c})\ \)for continuous features, \((d,z_{d})\) for discrete features, and \emph{save} or \emph{reset} for buttons.
\item
  \(T(s,\ a,\ s')\), which denotes the state-transition function and the environment was deterministic (\emph{i.e.}, \(P(s'|s,a) = 1\) for all \(s',s,a\)).
\item
  \(R(s,g,p)\), which provides the reward for state \(s\) under goal \(g\) for participant \(p\).
\end{enumerate}

We defined the participants' chair design decision strategy as \(\pi(a|s,g),\ \)where their next design action is conditioned on the current chair design and the goal. Every action leads to a state transition where the chair design and reward score are updated in real-time. 

\begin{figure}
    \centering
    \includegraphics[width=\linewidth]{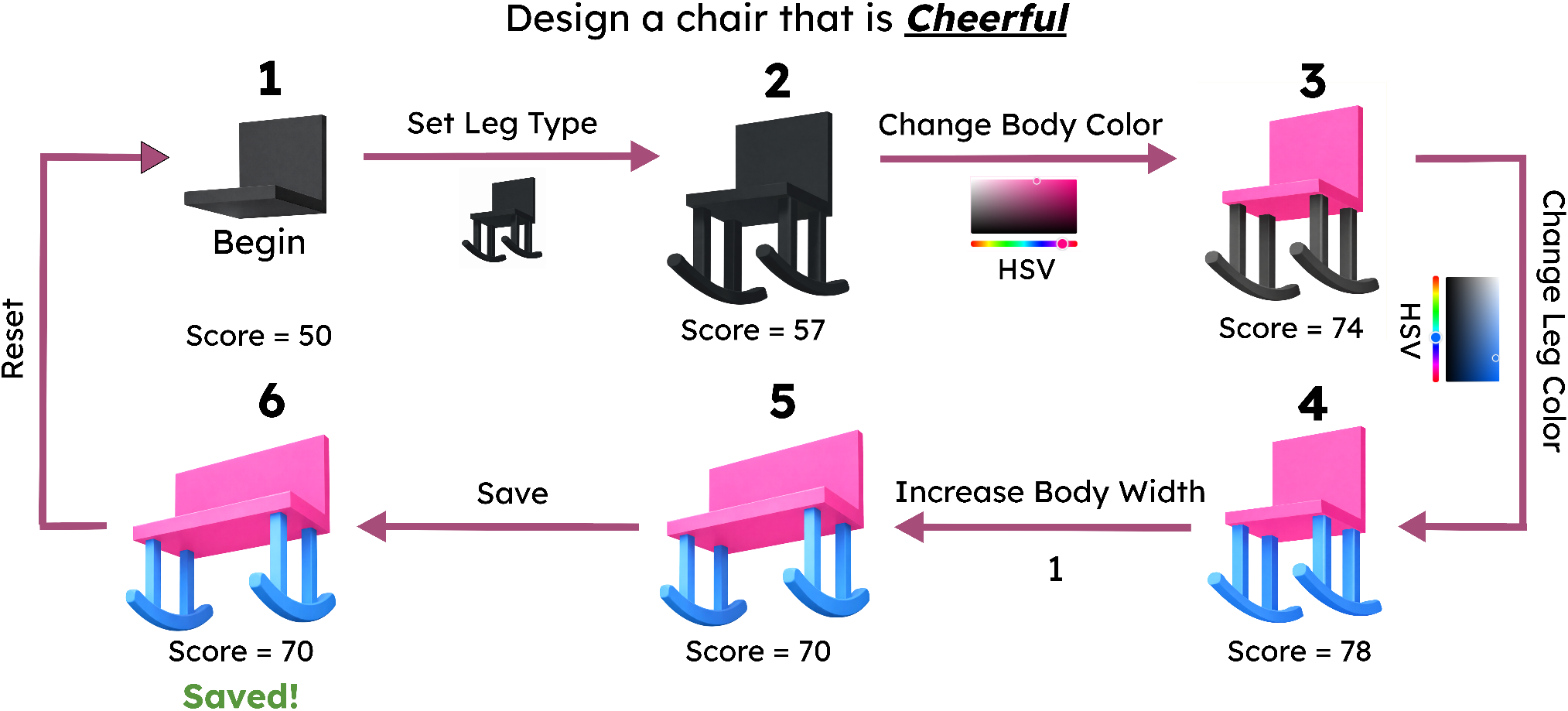}
    \caption{A sample action sequence, corresponding states, and rewards in the design MDP.}
    \label{fig:design-mdp}
\end{figure}

\autoref{fig:design-mdp} shows an example of chair design in the MDP framework.

\subsection{Goal-aligned Reward Signal}

We assumed that participants are naturally guided by an internal, intrinsic reward and our objective was to augment this reward with an extrinsic reward. Therefore, we decomposed \(R(s,g,p)\) = \(F(R^{int}(s,g,p),\ R^{ext}(s,g,p))\), (where \(R^{int}(s,g,p)\) is the intrinsic and \(R^{ext}(s,g,p)\) is the extrinsic reward component). Our contribution is designing \(R^{ext}(s,g,p)\) for the chair design task. While the choice of reward in a design task can be multifaceted and nuanced, our work represents a first step by constructing a shaping reward function \cite{ng1999policy} that provides precise, goal-relevant guidance towards higher-value designs.

We operationalised \emph{value} as proximity in design space to goal-relevant designs produced by an independent participant pool. Intuitively, a chair is considered higher value for a given goal if it resembles designs frequently produced by other designers in response to the same goal prompt. Here, this operationalisation corresponds to a higher likelihood of a chair design under a goal-specific distribution learned from pilot data. To construct this reward signal, we (1) learned a distribution per feature per design goal using a pilot dataset (N=120), which contained 223, 221, and 206 designs for \emph{cheerful}, \emph{dependable}, and \emph{unique} design goals, respectively. From the trained model, we (2) computed a score by evaluating the likelihood for each design, and transforming it into an integer between 0 to 100. This score served to guide participants toward regions of the design space that reflected stronger consensus for goal-relevant designs.

\textbf{Learning the distributions per goal.} To learn a distribution per feature per goal, we modelled the continuous and discrete features using separate likelihood families:

\begin{enumerate}
\def\labelenumi{\arabic{enumi}.}
\item
  Each continuous feature, \(c\ \epsilon\ C\), was modelled as a Normal distribution with goal-specific parameters \(\mu_{c,g}\ \)and \(\sigma_{c,g}\) denoting the mean and standard deviation of the distribution: \(x_{c}\sim N(\mu_{c,g},\sigma_{c,g})\)
\item
  Each discrete feature, \(d\ \epsilon\ D\), was modelled as a Categorical distribution with one goal-specific parameter \(\theta_{d,g,i} \geq 0\) per option \(i\): \(z_{d}\sim\ Cat(\overrightarrow{\theta_{d,g}})\), \({\sum_{i\  = \ 1}^{{|O}_{d}|}\theta_{d,g,i} =}_{}1\)
\end{enumerate}

Our formulation of the reward signal relied on two assumptions. First, we assumed independence among features. While this assumption may appear restrictive given the natural symmetry of everyday chair designs, an empirical analysis of feature correlations revealed that such dependencies were generally weak (\emph{r} \textless{} 0.3) in the chair design task. We therefore adopted the independence assumption as a reasonable simplification. Second, we modelled all continuous features using Normal distributions. This assumption could be violated if the design space exhibited strong multimodality or systematic bias. However, with the exception of hue (circularity) and saturation (bimodality), the remaining features displayed largely symmetric, unimodal distributions. These empirical properties are well captured by Normal approximations, making that assumption appropriate for the majority of the feature space.

\textbf{Computing the score.} With the estimated parameters, we computed the joint likelihood for any chair \(s = (x_{1},\ ..x_{|C|},\ z_{1},\ ..z_{|D|})\) designed for goal \(g\) by multiplying each learned, per-feature distribution evaluated at its feature value: 

\(L(s,g) = \ \prod_{c = i}^{|C|}N(x_{c}{;\mu}_{c,g},\sigma_{c,g})\ .\ \prod_{d = j}^{|D|}Cat(z_{d};\overrightarrow{\theta_{d,g}})\)

We then mapped the log-likelihood to an integer score in {[}0,100{]} using min-max normalisation. This gave us our participant-agnostic, goal-aligned reward signal.

We used hierarchical Bayesian unsupervised learning \cite{lee2014bayesian} to jointly learn all parameters per goal. We trained the model in \href{https://www.pymc.io/welcome.html}{PyMC}, with (18 × 2) + (3 + 8 + 9) = 56 learnable parameters per goal. We ran NUTS \cite{hoffman2014nuts} (4 chains; 1000 warm-up + 1000 draws per chain), and used posterior means as fitted parameter estimates.

\subsection{Goal-agnostic reward signal}

To isolate effects driven by \emph{reward} feedback from effects driven by \emph{goal} feedback, we also introduced a participant-specific, goal-agnostic reward \(R_{goal - agnostic}(s,g,p)\). This reward signal matched the learned reward in structure, determinism, and complexity (it defines a smooth, optimisable landscape over the same features), but was intentionally goal-agnostic: for each participant \(p\) assigned to the goal-agnostic condition in the reward phase, we generated a random set of parameters:

\(\mu_{c,g,p}\), \(\sigma_{c,g,p}\)\(\ \forall\ c\ \epsilon\ C\) \textasciitilde{} {[}0,1{]}, 
and 

\(\overrightarrow{\theta_{d,g,p}\ }\) \(\forall\ d\ \epsilon\ D,\ \theta_{d,g,p,i}\ \sim\ \lbrack 0,1\rbrack,\ s.t. {\sum_{i\  = \ 1}^{|O_{d}|}\theta_{d,g,p,i} =}_{}1.\ \)

These parameters defined a participant-specific likelihood function \(L(s,g,p)\) of the same form as that of the goal-aligned reward (Eqn 3). We then applied the same min-max normalisation to obtain the score. As a result, participants received a deterministic, stepwise score that could be optimised. The direction of optimisation, however, was arbitrary and unrelated to the goal.

\section{Behavioural Study}

\subsection{Participants} 
353 participants were recruited online via Prolific (172 females; mean age = 40.8 \(\pm 11.8\) years). Participants were compensated at 17 USD per hour. The study was approved by Toyota Research Institute's Institutional Review Board.

\subsection{Procedure}
In our study, each participant completed three design phases of five minutes each. First was a \textbf{practice phase} with an intrinsic goal to ``design chairs you like''. This phase was intended for participants to freely explore the design environment, become familiar with the features, and design chairs purely based on their idiosyncratic preferences. Second was a goal-driven creation phase (\textbf{baseline phase}) where participants were given one of the three abstract extrinsic design goals---\emph{cheerful}, \emph{dependable} or \emph{unique}---and were instructed to design chairs that reflected the goal (``design chairs that are \textless goal\textgreater''). The third and final phase was a reward-guided design phase (\textbf{reward phase}), where participants continued to design chairs for the same goal as in the baseline phase, except they now received a score with every action (``design chairs that are \textless goal\textgreater. You will receive a score at each step.''). Participants received one of two reward types: a \textbf{goal-aligned} reward signal or a participant-specific, \textbf{goal-agnostic} reward signal, and were free to use the score however they pleased (``how you choose to use the score is up to you''). The between-participants study design was therefore 3 × 2 (conditions: goal × reward type). In each phase, participants were instructed to design at least two chairs and remain active. 

\begin{figure}
    \centering
    \includegraphics[width=\linewidth]{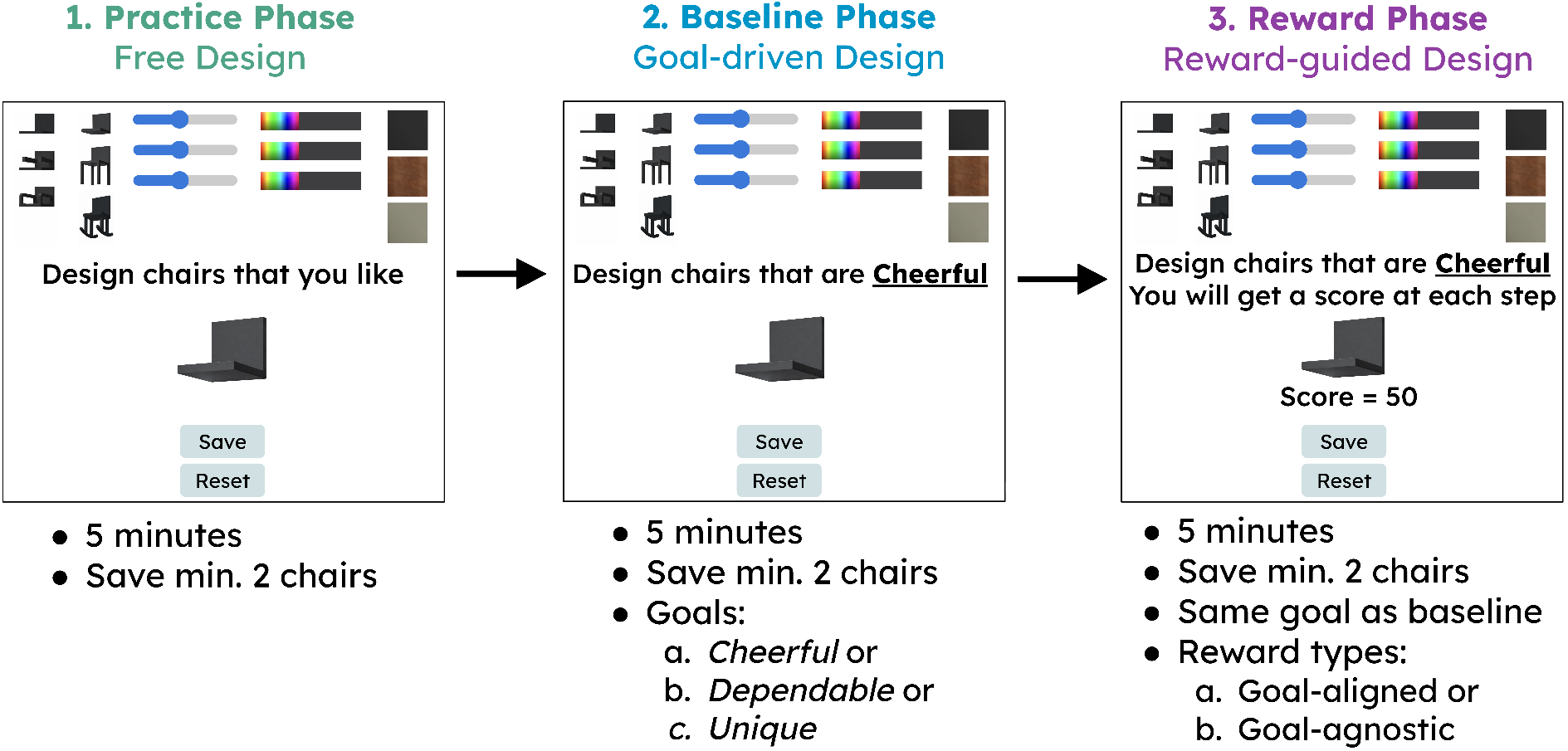}
    \caption{Experimental procedure comprising three phases: practice, baseline (with only a goal) and reward (with both a goal and reward feedback).}
    \label{fig:experimental-procedure}
\end{figure}

\autoref{fig:experimental-procedure} illustrates the full study pipeline with the three phases.

After each phase, participants were asked about their experience and could opt to take a break before proceeding to the next phase. Participants who did not save at least two chairs within five minutes received a warning and a two-minute extension; failure to meet this requirement within seven minutes resulted in the study timing out and an end to their participation (no timer was visible). After completing the three phases, participants were asked about their subjective experience, including decision making strategies and interaction with the interface (adapted from the Creativity Support Index, CSI \cite{cherry2014csi}); participants also provided demographics and unstructured comments. Overall, participation averaged approximately 40 minutes (mean duration = \(40.3 \pm 12.7\) minutes). \autoref{tab:participant-conditions} shows the number of participants per condition.

\begin{table}[htbp]
\centering
\caption{Participant conditions (Total = 353).}
\label{tab:participant-conditions}
\begin{tabular*}{\textwidth}{@{\extracolsep{\fill}}lccc}
\toprule
 & \emph{Cheerful} & \emph{Dependable} & \emph{Unique} \\
\midrule
Goal-aligned & 59 & 60 & 59 \\
Goal-agnostic & 57 & 60 & 58 \\
\bottomrule
\end{tabular*}
\end{table}

\section{\\Results}

\subsection{Data Quality Assessment}

\begin{figure}
    \centering
    \includegraphics[width=\linewidth]{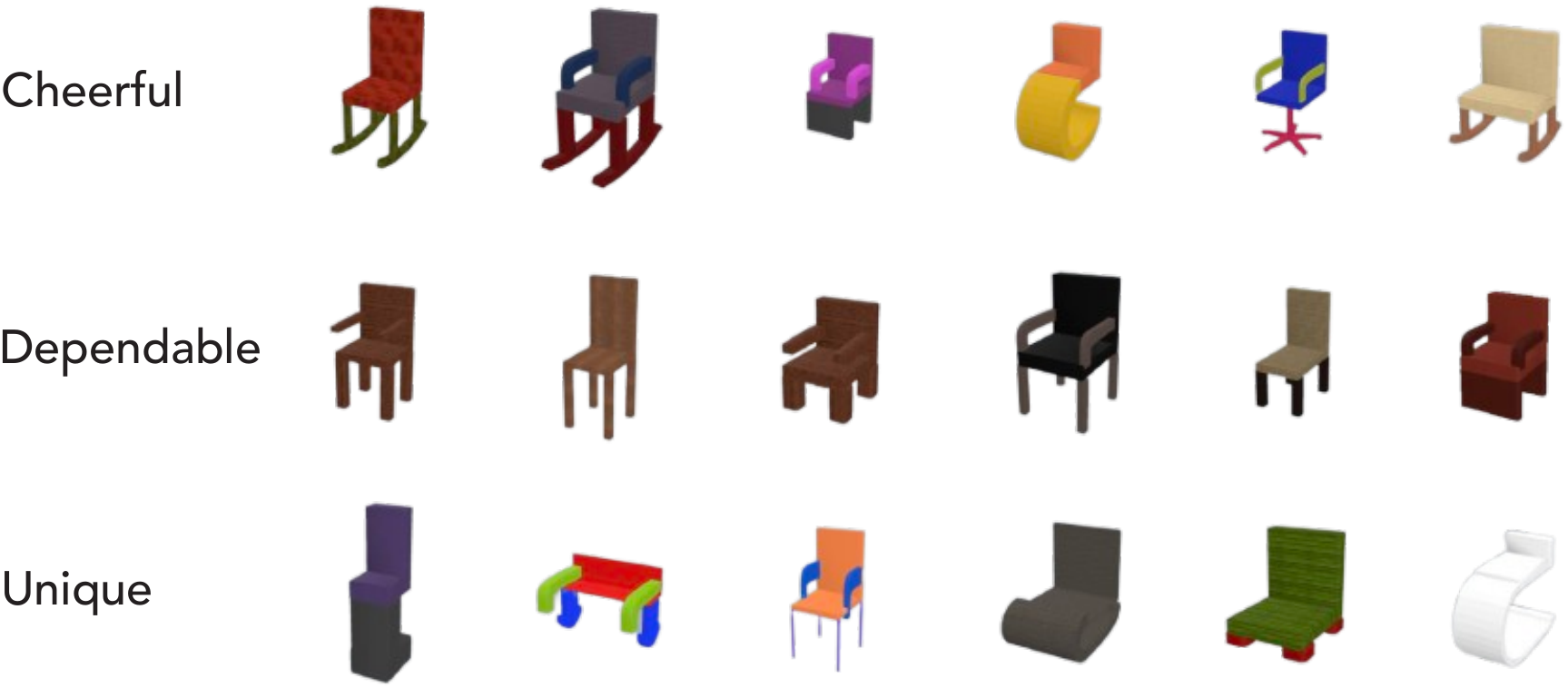}
    \caption{Example designs by goal designed by study participants in the baseline phase.}
    \label{fig:example-designs}
\end{figure}

We assessed the quality of the study data and participant engagement. Participants rated the task interface highly in supporting their creativity, expressiveness, and satisfaction, with average scores of 8.45, 8.33, and 8.23 (of 10), respectively. Engagement was also reflected in participant behaviour: nearly 70\% of participants saved more than the minimum requirement of two chairs per phase. Moreover, nearly 25\% of participants used an optional free response box to describe the study as ``fun'', ``engaging'', or ``creative''. Example chair designs from the baseline phase per goal are shown in \autoref{fig:example-designs}. Chairs were highly diverse and reflected interesting goal-specific patterns---\emph{cheerful} was associated with colour and movement, \emph{dependable} with sturdiness and durability, and \emph{unique} with atypical configurations.

\subsection{Reward Signal Validation}

\begin{figure}
    \centering
    \includegraphics[width=\linewidth]{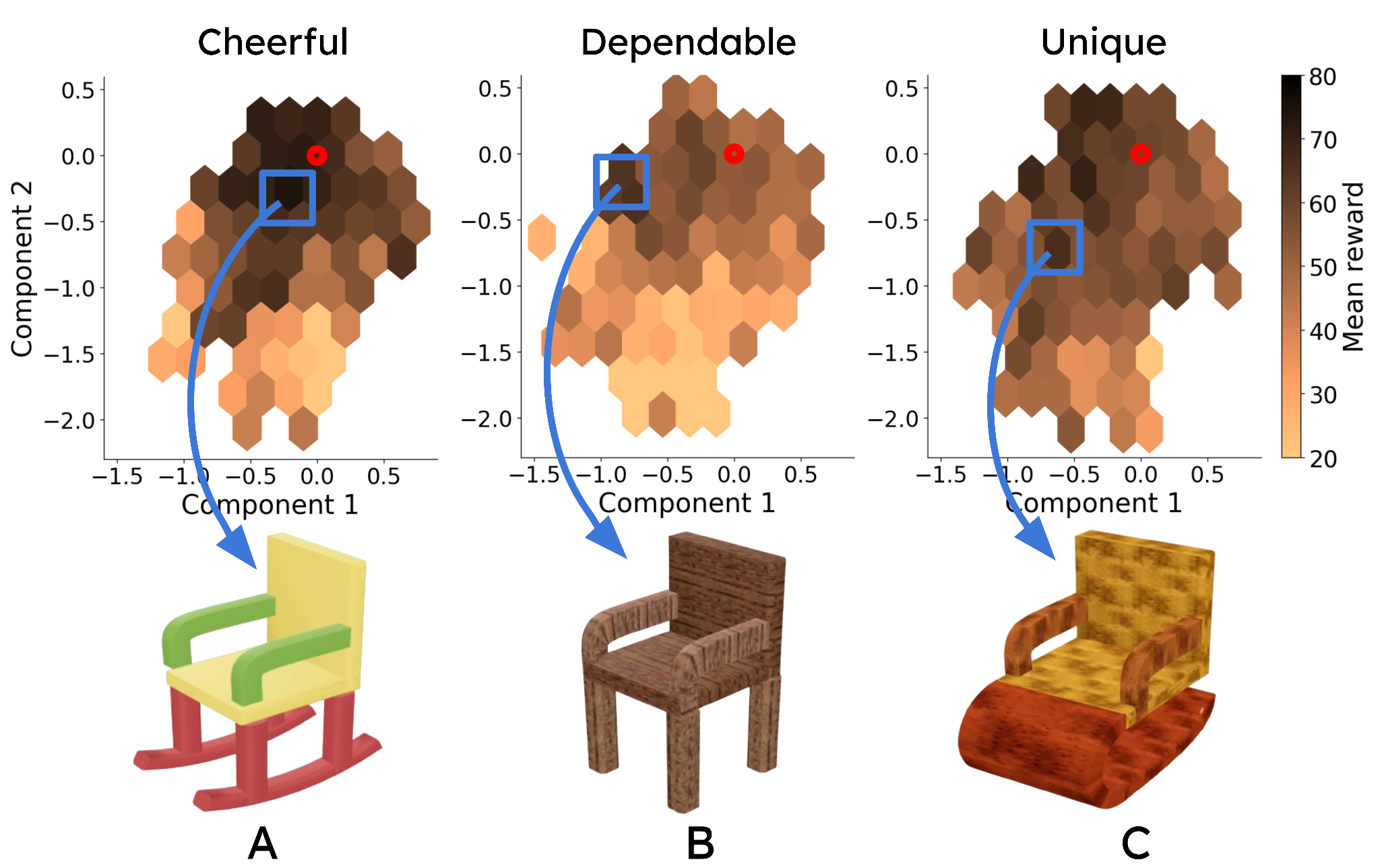}
    \caption{Learned reward landscape with an example high-scoring design per goal. The starting configuration is marked by the red dot and high reward regions are marked by a blue square.}
    \label{fig:reward-landscape}
\end{figure}

We validated the structure of the goal-aligned reward signal for each goal by visualising the learnt reward landscape following a principal component analysis of design space. To represent our design space, we trained an autoencoder on the saved chairs from the practice phase and used the latent embedding space as the design representation space. We then visualised the learnt reward landscape for each goal by projecting all saved designs onto a shared low-dimensional space using Truncated SVD \cite{hoffman2014nuts}. We discretised this space into bins and computed the mean learned score within each bin. \autoref{fig:reward-landscape} shows the learnt reward landscapes with an example high-scoring design for each goal. The red circle marks the starting configuration, and the blue square indicates regions associated with high reward for the corresponding goal. The \emph{unique} learnt landscape is more diffuse than for cheerful and dependable, indicating greater spread in the learned feature distributions. The differentiated locations of these high-reward regions across goals validates the approach.

The goal-agnostic reward, in contrast, yielded a similar structured and deterministic landscape as the goal-aligned reward. Importantly, its gradient was oriented along an arbitrary direction in design space that varied by participant and was unrelated to the goal. The correlation between goal-aligned and goal-agnostic rewards across participants was negligible (\emph{r} = 0.05), establishing the goal-agnostic reward as a goal-irrelevant, matched control for the presence of optimisable feedback. Taken together, these two reward types allowed us to disentangle the effects of reward availability and quality on goal-conditioned design decision making.

\subsection{Impact of Rewards on Design Decision}

The structure of the MDP allowed investigation of how reward shapes the design process and its outcomes. We examined the action space (interaction patterns), reward space (score optimisation), and design space (chair representations).

\begin{figure}
    \centering
    \includegraphics[width=0.9\linewidth]{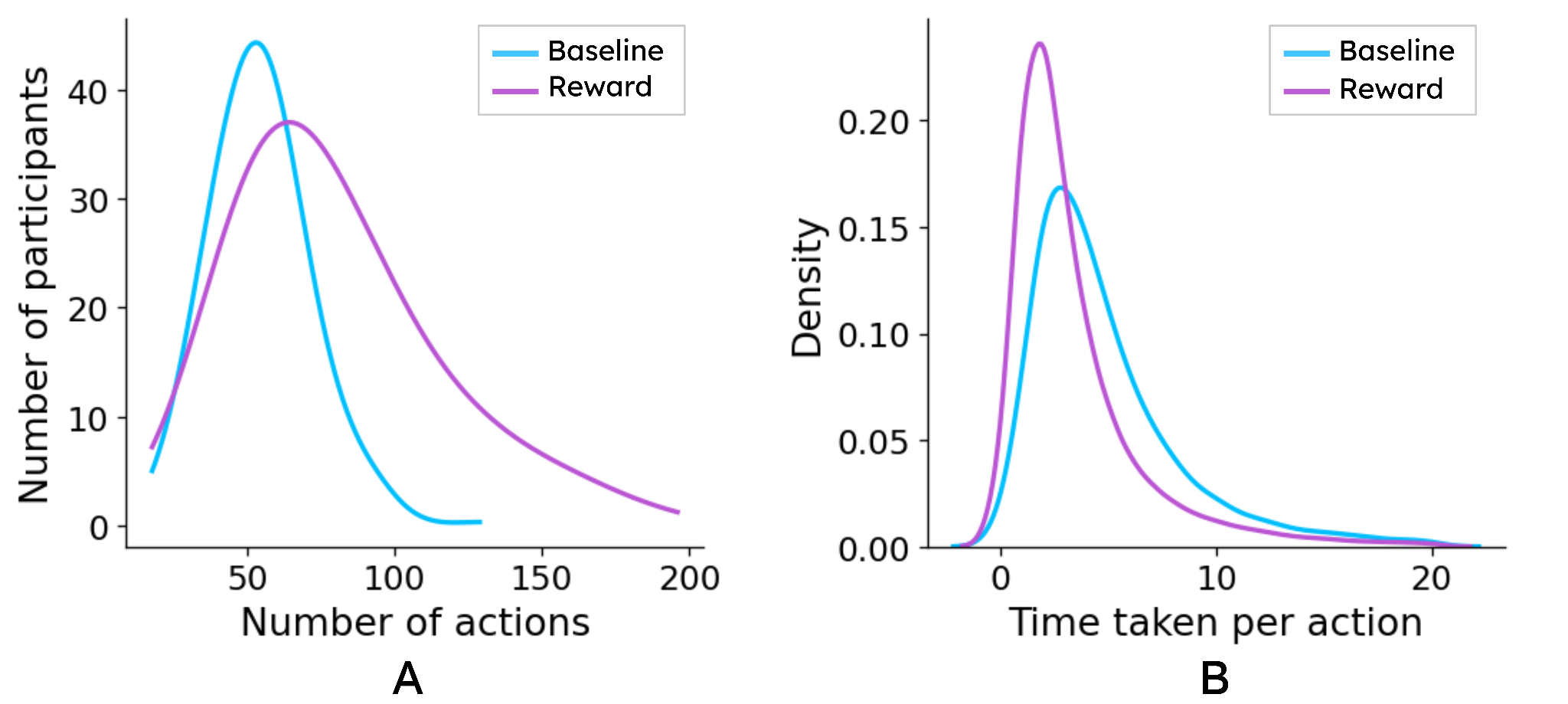}
    \includegraphics[width=0.9\linewidth]{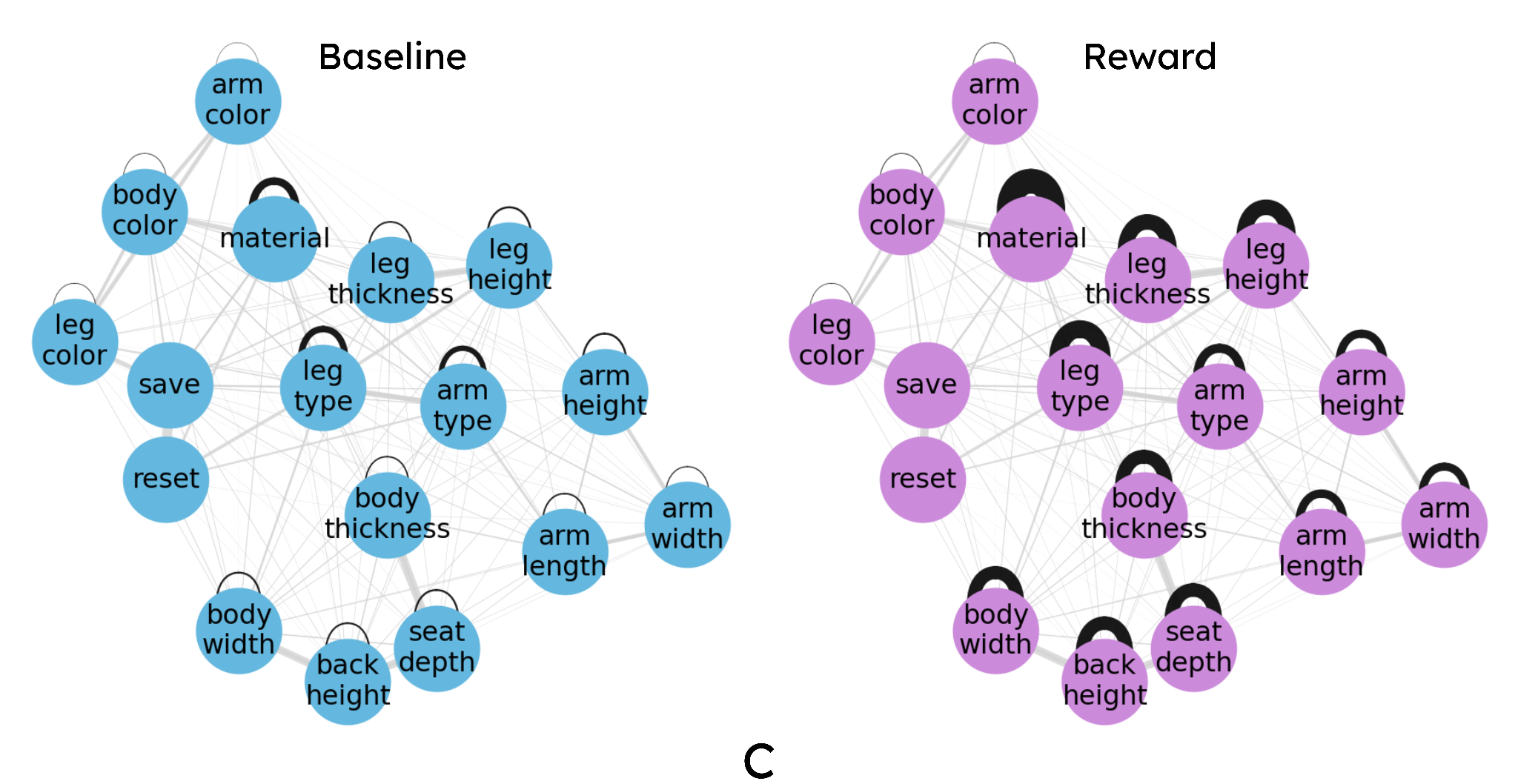}
    \caption{Effect of reward feedback on action space. (A) Distribution of number of actions in the baseline and reward phases. (B) Distribution of the time taken per action in the baseline and reward phases. (C) Markov action transition graphs for the baseline and reward phases.}
    \label{fig:action-space}
\end{figure}

\textbf{Reward feedback impacted the design process.} We tested how stepwise rewards impacted the design process by considering individual participant actions (\autoref{fig:action-space}). A linear mixed-effects analysis revealed that participants performed significantly more actions in the reward phase than in the baseline phase ($\beta$ = 25.05, SE = 1.61, z = 15.59, p \textless{} .001, 95\% CI {[}21.90, 28.20{]}), suggesting increased engagement under reward feedback (\autoref{fig:action-space}A). Further, relative to baseline, the time taken per action decreased in the reward phase ($\beta$ = -1.56, SE = 0.10, z = -15.08, p \textless{} .001, 95\% CI {[}-1.76, -1.36{]}), indicating that people made quicker actions under reward feedback (\autoref{fig:action-space}B). 

We also analysed how feedback changed the action policy by constructing a Markov transition graph over action types for each phase (\autoref{fig:action-space}C). Compared to baseline, the reward phase exhibited higher persistence in features (t(352) = 15.85, p \textless{} .001), suggesting that reward feedback altered the quantity and speed of interaction, as well as the structure of participants' design strategies.

\begin{figure}
    \centering
    \includegraphics[width=\linewidth]{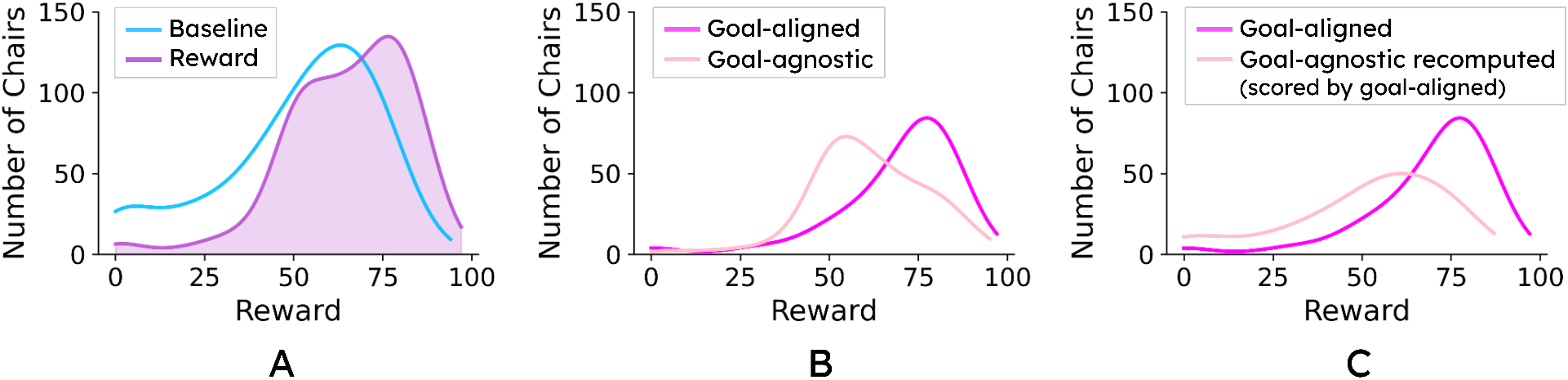}
    \caption{Reward maximisation by goal condition. (A) Distribution of rewards for baseline (computed post-hoc using goal-aligned reward signal) and reward phase (attained by the respective reward signal). (B) Decomposition of reward distribution by goal condition (goal-aligned vs goal-agnostic) in the reward phase. (C) Recomputing the reward distribution under goal-agnostic reward type when scored with respect to the goal-aligned reward signal.}
    \label{fig:reward-maximisation}
\end{figure}

\textbf{All rewards steered behaviour, but goal-aligned was maximised more than goal-agnostic feedback.} We used the goal-aligned reward signal to assign scores \emph{post hoc} to all chairs saved in the baseline phase (where participants designed for a goal but received no feedback). We then compared these scores to those of chairs saved in the reward phase (based on the respective goal-aligned or goal-agnostic reward signal). Compared to baseline, rewards were distributed differently (\emph{i.e.}, bimodally rather than unimodally) (\autoref{fig:reward-maximisation}A). Using mixed-effects modelling, we found that reward type explained this bimodality: chairs designed under goal-aligned rewards scored significantly higher than chairs designed under goal-agnostic rewards ($\beta$ = 8.40, SE = 1.52, z = -5.51, p \textless{} .001, 95\% CI {[}-11.39, -5.41{]}) (shown in \autoref{fig:reward-maximisation}B).

Further, for the chairs designed under goal-agnostic rewards, we recomputed their reward scores using the \emph{goal-aligned} reward signal (instead of the \emph{goal-agnostic} reward signal) to test participants' goal-optimization behaviour under misaligned rewards. Rescoring showed a rightward shift in the score distribution (t(174) = 6.45, p \textless{} .001, \autoref{fig:reward-maximisation}C), suggesting that, even when presented with a goal-agnostic reward signal, participants instead continued to optimise their designs for the instructed task goal, and were not misled by the goal-agnostic reward.

To quantify the effect of rewards at the participant level, we computed the \emph{positive reward drift}, or the difference between a participant's mean reward in the reward phase and their mean reward in the baseline phase. Mean drift was 19.12 under the goal-aligned reward feedback and 9.64 under the goal-agnostic reward (t = 4.09, p \textless{} .001), indicating substantially stronger steering when reward feedback was goal-relevant. To further interrogate how the goal-relevance impacted behaviour, we examined instances wherein the goal-agnostic reward yielded stronger improvements. We found marginal evidence that participants attained higher mean scores in the reward phase when their participant-specific, goal-agnostic reward feedback optimum was closer to the goal-aligned reward feedback optimum (r(173) = -.14, p = .07). Thus, when goal-agnostic rewards were incidentally aligned with the goal-aligned reward landscape, participants maximise them (marginally) more. This suggests that while the presence of a score impacts behaviours, the magnitude of impact is driven by the quality of the feedback---\emph{i.e}., how goal-relevant the score is.

\begin{figure}
    \centering
    \includegraphics[width=\linewidth]{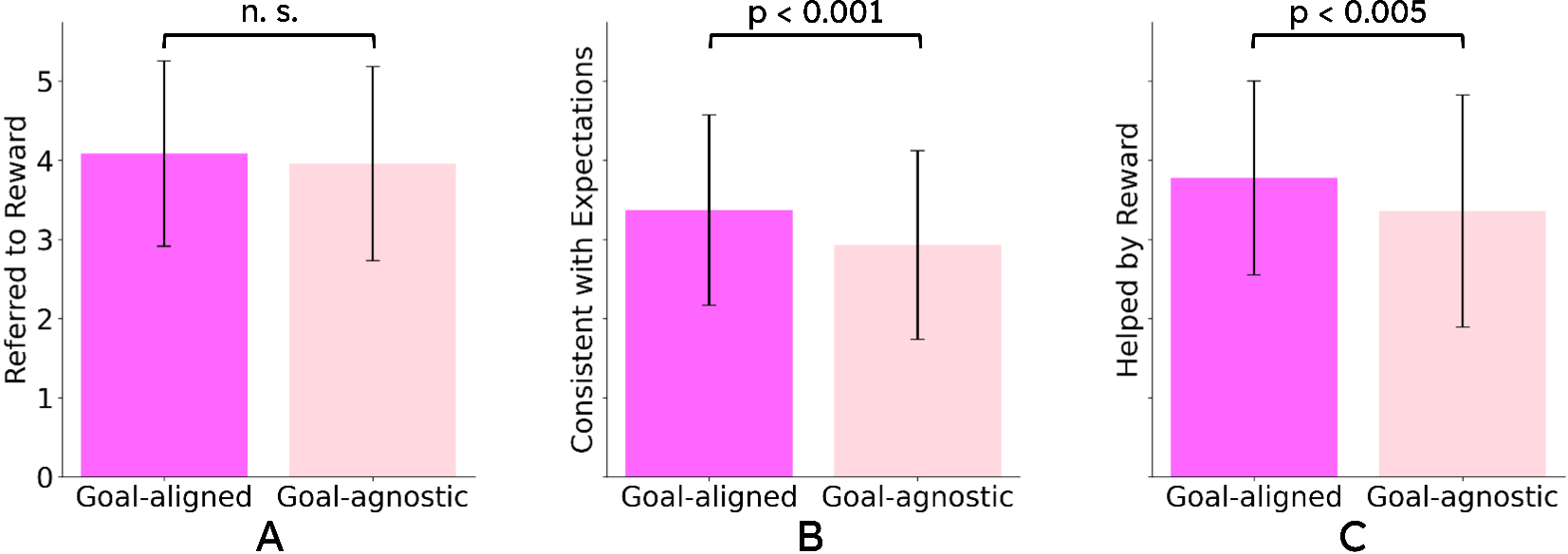}
    \caption{Participant ratings by reward-type (1-5) for (A) how much they referred to the score, (B) how consistent the score was with their expectations, and (C) how useful the score was for designing goal-relevant chairs.}
    \label{fig:participant-ratings}
\end{figure}

Participants' subjective experience complemented these results. \autoref{fig:participant-ratings} shows participant ratings on a five-point, Likert-style scale for how much they referred to the reward (\autoref{fig:participant-ratings}A), how consistent the reward was with their expectations (\autoref{fig:participant-ratings}B), and how helpful they perceived the reward to be for designing goal-relevant chairs (\autoref{fig:participant-ratings}C). Participants did not differ significantly in how much they referred to the reward across reward types (t = 0.98, p = .33). In contrast, goal-aligned rewards were rated as significantly more consistent with participants' expectations than goal-agnostic rewards (t = 3.45, p \textless{} .001), and were also perceived as significantly more helpful for designing goal-relevant chairs (t = 2.89, p \textless{} .005).

More generally, we analyzed participants' unstructured comments on how the score helped and harmed them in the task. The analysis revealed nuanced insights into how participants responded to the score. The score was found to be helpful as guidance to improve designs, enabling participants to explore and learn about the utility of different controls for achieving the goal, and boost confidence in their choices. On the other hand, the reward harmed participants when it was perceived as a distraction that induced anxiety or reduced their sense of agency. There was also a split between some participants, who reportedly either gamified the interaction, explicitly trying to maximise the score, or completely ignored the score in their design decisions.

\textbf{Goal-aligned feedback moved designs toward more optimal regions.} The impact of feedback in the reward space was mirrored in the design space. For each saved chair, we computed its similarity to the optimal chair as (1 $-$ \emph{Gower distance}), where the optimal chair was defined as the highest likelihood design under the relevant reward signal (\emph{i.e.}, goal-aligned reward signal for baseline phase, and either goal-aligned or goal-agnostic reward signal for reward phase). Consistent with findings in the reward space, chairs saved in the reward phase exhibited a bimodality with respect to their similarity to the optimal chair that could be explained by the reward type ($\beta$ = $-$0.154, SE = 0.006, z = $-$26.24, p \textless{} .001, 95\% CI {[}$-$0.165, $-$0.142{]}). These findings indicated that goal-aligned feedback not only increased reward scores, but also steered designs toward more optimal regions of the design space.

\begin{figure}
    \centering
    \includegraphics[width=\linewidth]{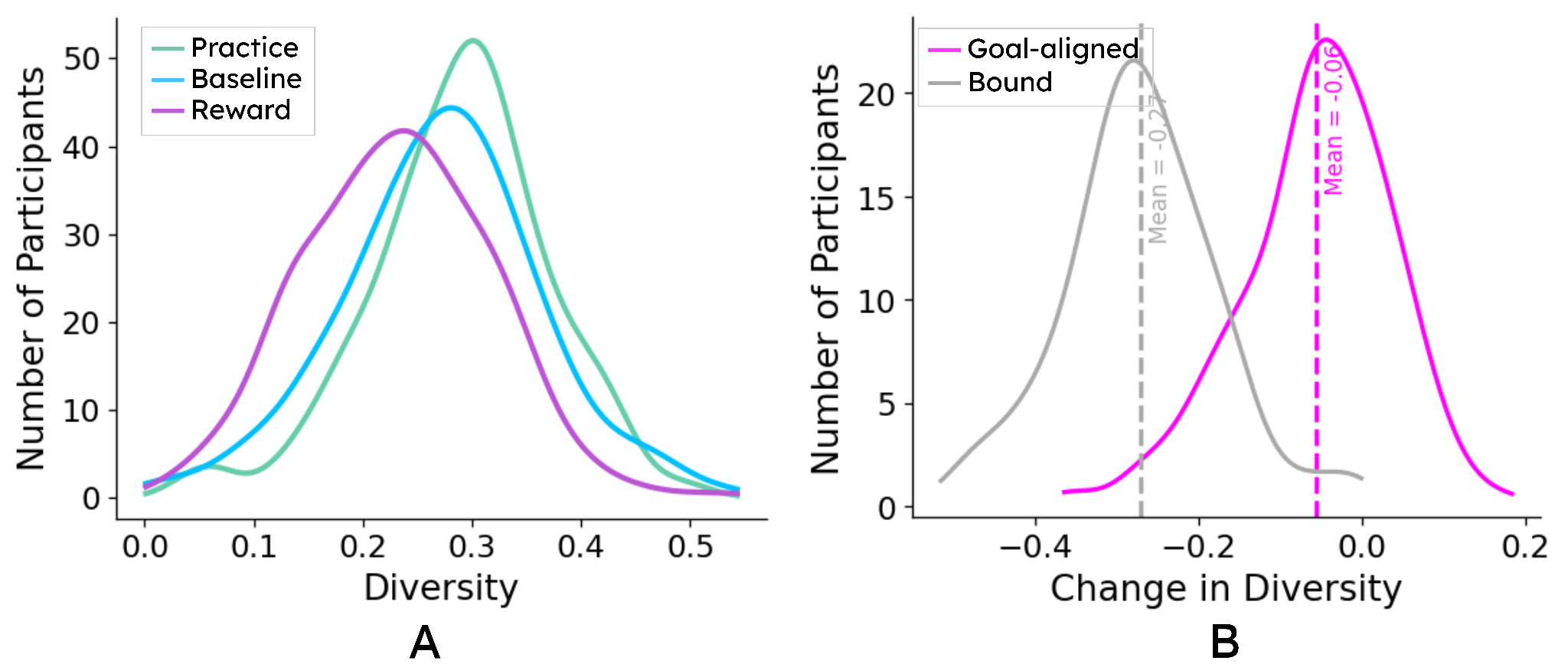}
    \caption{(A) Distributions of design diversity in practice, baseline, and reward phases. (B) Distribution of change in diversity from baseline to reward phase for the goal-aligned reward. The grey distribution marks the upper bound of the theoretically possible diversity drop.}
    \label{fig:design-diversity}
\end{figure}

\textbf{Design diversity reduced with increasing constraints, but meaningful exploration remained.} We then tested whether reward steering is associated with reduced design diversity. We quantified participant-level diversity as the mean pairwise distance between the set of designs each participant saved within a phase. We observed that participant-level diversity decreased systematically across phases as the task became increasingly constrained: from practice to baseline phase ($\beta$ = $-$0.018, p \textless{} .001), and then from the baseline to the reward phase ($\beta$ = $-$0.057, p \textless{} .001) (\autoref{fig:design-diversity}A). For the goal-aligned reward, we also computed the change in diversity across phases (reward $-$ baseline) and found a significant reduction in diversity (W = 3113, p \textless{} .001, median $\Delta$ = $-$0.048). 

To contextualise the magnitude of this reduction, we estimated an upper bound on the change in diversity by considering the extreme case in which all designs converge to the optimal design in the reward phase. This estimated bound represented the largest possible diversity drop due to reward steering. Importantly, the observed reduction in diversity was significantly smaller than this theoretical maximum (W = 15931, p \textless{} .001, median $\Delta$ = 0.220). Instead, the reduction was closer to zero, suggesting that rewards steered designs toward higher-value regions of the reward landscape without collapsing exploration entirely (\autoref{fig:design-diversity}B).

\textbf{Reward impact varied by goal condition---\emph{unique} is unique.} We further examined whether rewards obtained during the reward phase differed by specific goal. Across conditions, we found no significant difference in final score between goals (\emph{dependable} vs. \emph{cheerful}: $\beta$ = $-$2.88, SE = 1.94, z = $-$1.49, p = .14; \emph{unique} vs. \emph{cheerful}: $\beta$ = $-$0.55, SE = 1.95, z = $-$0.28, p = .78), nor any significant interaction effects between goal and reward type (\emph{dependable}: ($\beta$ = 6.07, SE = 3.71, z = 1.64, p = .10), \emph{unique}: ($\beta$ = 6.18, SE = 3.74, z = 1.66, p = .10)). Nonetheless, we observed qualitative effects of goal on the design scores and the self-perception of rewards.

\begin{figure}
    \centering
    \includegraphics[width=\linewidth]{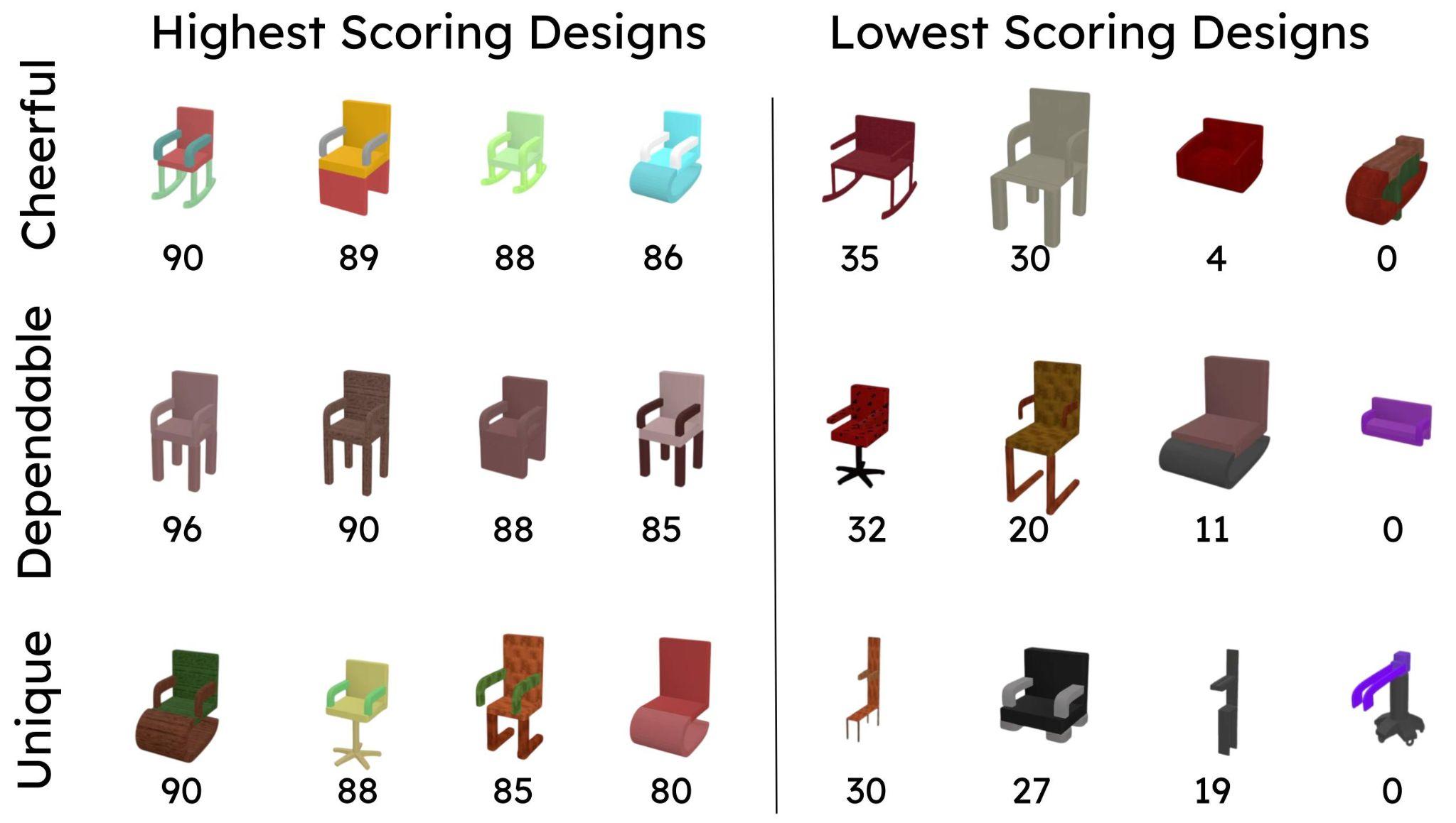}
    \caption{Example high- and low-scoring designs for the goal-aligned reward.}
    \label{fig:high-low-designs}
\end{figure}

To qualitatively interpret the saved designs, we visualised representative high- and low-scoring designs for each goal under the goal-aligned reward (\autoref{fig:high-low-designs}). High-scoring chairs matched interpretable, goal-consistent characteristics (e.g., \emph{cheerful} associated with color and movement; \emph{dependable} with sturdiness and durability) yet still showed substantial variation. Low-scoring chairs deviated from goal-consistent characteristics. The \emph{unique} goal, however, was a notable exception to this pattern. Unlike \emph{cheerful} and \emph{dependable}, the goal-aligned reward for \emph{unique} appeared less effective at capturing the value of atypical configurations for the \emph{unique} goal. The distribution of \emph{unique} designs was less consensual and more multimodal, likely because \emph{unique} chairs could be unusual in multiple distinct ways. As a result, a likelihood-based, consensus-seeking reward may be a weaker proxy for \emph{uniqueness}, as also reflected in a flatter learned reward landscape (shown in \autoref{fig:reward-landscape}C).

This interpretation was aligned with participants' self reports. When we stratified ratings of score usage (\autoref{fig:participant-ratings}A), consistency with expectations (\autoref{fig:participant-ratings}B), and perceived usefulness (\autoref{fig:participant-ratings}C) by goal, we observed a clear goal-dependent pattern. Across all three goals, participants reported referring to the score to a similar extent under goal-aligned and goal-agnostic feedback, suggesting comparable attention to the signal. However, for \emph{cheerful} and \emph{dependable}, goal-aligned feedback was rated as significantly more consistent with participants' expectations and more useful for designing goal-relevant chairs. By contrast, for \emph{unique} designs, goal-aligned and goal-agnostic feedback did not differ significantly in perceived consistency (t = 0.72, p = .47) or usefulness (t = $-$0.56, p = .58), indicating that participants did not perceive the goal-aligned reward as a reliable or useful signal for achieving uniqueness.

In sum, our findings indicate that the impact of reward feedback depends on both the quality of the signal and the nature of the goal: consensus-based goal-aligned rewards drive larger improvements than matched goal-agnostic feedback, yet goals such as \emph{unique} may require differently designed rewards to capture goal attainment.

\section{Discussion}

Here, we present a systematic investigation of the role of reward feedback in design decision making. We modelled a feature-rich, goal-conditioned chair design task as an MDP. We designed two reward signals of identical structure and complexity and used them to compute scores that could be presented to participants at every step of the design process. Half of the participants received (1) a goal-aligned reward signal by learning a distribution on goal-relevant chairs while the other half received (2) a participant-specific, goal-agnostic reward signal with arbitrary reward gradients. We then leveraged the power of the MDP framework to conduct a multi-level analysis of feedback effects across action, reward, and design spaces. Participants' behaviour and subjective experience was assessed for three goals: \emph{cheerful}, \emph{dependable}, and \emph{unique}.

Our results demonstrated that the presence of reward feedback altered the nature of participant actions during the design process. When provided with reward feedback, participants made more actions more quickly, suggesting increased interaction vigor. We speculate that this invigoration is consistent with computational accounts in which higher reward rates increase the opportunity cost of time, and hence response vigor \cite{niv2007dopamine}. Reward also altered the \emph{structure} of action selection, with participants exhibiting higher persistence on individual features. This may reflect a shift from \emph{diversive} to \emph{specific} exploration \cite{berlyne1960conflict}\emph{---i.e.}, rather than broadly sampling across feature controls, participants adjusted a single feature for longer, perhaps to refine local hypotheses about how each change would impact the score. In Schön's terms, the score changed the nature of ``back-talk'' from the design in its ongoing dialogue with the designer, encouraging rapid tweak--evaluate loops that support design optimisation \cite{schon1992designing}.

Importantly, our results showed that participants consistently increased the goal-aligned but not the goal-agnostic score, suggesting that extrinsic rewards were not accepted indiscriminately. Instead, participants appeared to maintain an intrinsic goal-consistency criterion (a subjective evaluation of goal attainment) and optimised external scores primarily when they were perceived as informative for maintaining that criterion. Otherwise, they resorted to other signals such as intrinsic rewards to guide goal attainment without being misled by low-quality signals. In other words, although rewards shaped behaviour as predicted, they did not seem to shape design intent; instead, rewards were applied \emph{in service of} intent, especially when they were goal-aligned. Such behaviour is consistent with the Situated Function-Behaviour-Structure framework \cite{gero2004situated}, wherein design is considered to be a sequence of transformations between function, expected behaviour, and structure, with an \emph{evaluation} step comparing derived and expected behaviour. Under this framework, designers continuously test whether a structure ``behaves'' as intended relative to their goal representation, rather than blindly accepting an external score. These results are also in line with metacognitive control accounts of creativity \cite{lebuda2023metacognition}, which suggest that monitoring processes evaluate candidate actions and outcomes, while control processes regulate strategy selection. Thus, we identified that \emph{useful} rewards were optimised, while being robust to ``reward hacking'' \cite{pan2022reward}.

Unsurprisingly, design under the influence of rewards was accompanied by a reduction in design diversity. However, this reduction was small compared to the estimated maximum possible effect, and qualitative inspection revealed that variation across designs remained high. Thus, although reward feedback steered designs toward optimal regions in design space, diversity did not collapse entirely. A possible reason may be the high dimensionality of the design space; in this high-dimensional space, designs can be near-optimal along several axes, resulting in sufficient variation across designs. The preservation of design diversity under meaningful reward feedback is particularly relevant in the context of AI-mediated creativity support, where AI feedback has been linked to homogeneity \cite{anderson2024homogenization}. Nonetheless, convergence in the design space is also valuable, as it helps refine ideas and better satisfy constraints. Because rewards can be applied flexibly, they can be designed to balance both divergent and convergent thinking in creative design \cite{cross2021engineering,liu2003concept,nath2024creative}. This way, they can support different phases of the process and also adapt to the context and individual.

We found no statistically significant effects of goal type on reward optimisation behaviour. Thus, although goal identity is important and is known to influence the design process and its outcomes \cite{nandy2025semantic}, its effects may be difficult to detect in aggregate behavioural measures due to the high heterogeneity of design outcomes, and may be mediated by other mechanisms \cite{gollwitzer1996goal,busch2023decoding}. However, goal type significantly impacted \emph{self-perception} of the rewards. The goal-aligned reward signal was perceived as more useful than the goal-agnostic reward signal for \emph{cheerful} and \emph{dependable}, but not for \emph{unique}. \emph{Cheerful} and \emph{dependable} goals had higher levels of consensus in design outcomes, with interpretable feature patterns. In such cases, a likelihood-based reward trained on prior examples was perceived as a useful proxy for goal attainment. In contrast, \emph{unique} is an inherently relational concept, defined by its distinction from what is typical, yielding a multimodal distribution. It is therefore reasonable that a consensus-seeking likelihood reward would be perceived as less useful, as it undervalues genuine uniqueness. Such goal-specific effects encourage consideration of other forms of reward signals. For example, quality-diversity \cite{pugh2016quality} based rewards, which recognise high-value solutions spread across diverse regions of design space, could be well-suited for goals such as \emph{unique}. Overall, these findings underscore the relative importance of the \emph{degree of goal alignment} over the specific semantic meaning of a goal when designing rewards.

\subsection{Limitations}

While this work reveals aspects of how rewards shape goal-driven design behaviour, it has noteworthy limitations. Our goal-aligned reward depended on pre-existing data, and its modelling assumptions may have been ill-suited to address feature correlations, circularity, or multi-modality (\emph{e.g.}, in the case of colour). The study phases were not randomised, making order effects difficult to disentangle. Additionally, substantial heterogeneity across participants and their chair designs may have affected our ability to detect main or higher-order effects of goal type in a sample of this size. Finally, while our task had many features of commonly used parametric design tools, the interface and incentive structure were less complex than real-world design practice, limiting their ecological validity.

\subsection{Future Work}

Our work represents only a first step toward understanding how feedback influences design thinking. The MDP formalism of parametric design provides process-level insights into design and can be used to test hypotheses at multiple levels of abstraction which can drive the construction of precise rewards for targeted behaviour steering. Building on our findings, we outline a forward-looking guide for holistic understanding of the impact of rewards in design decision making, and for the future of facilitating effective design thinking without compromising designer agency or creative well-being \cite{ryan2000self}. Reward design should consider:

\begin{itemize}
\item
  \textbf{What:} \emph{What behaviours should be rewarded?} Our goal-aligned reward operationalises goal consistency via a consensus proxy, which proved effective for goals with shared semantic structure but less so for potentially multimodal goals (\emph{unique}). Future rewards can target alternative constructs such as exploration, creativity, or multi-objectives such as quality--diversity \cite{pugh2016quality}.
\item
  \textbf{How:} \emph{How should the reward be presented?} Instead of presenting rewards as a scalar score, future rewards can leverage richer forms, including those already common in modern AI-assisted tools such as candidate designs, contrastive examples, natural-language critiques, or even affective cues. Such interfaces may preserve designer agency by making feedback more interpretable, supporting reflection and active choices about which feedback to incorporate.
\item
  \textbf{When:} \emph{When should the reward be presented?} Our results suggested that the usefulness of rewards may be phase-dependent, with convergence toward higher-value regions providing particular benefit during late-stage refinement. Adaptive rewards, with varying reward signals across stages, are promising (\emph{e.g.}, early rewards for novelty and late rewards for goal-consistency \cite{nath2024creative}).
\item
  \textbf{Whom:} \emph{To whom is the reward being presented?} Another important direction concerns individual differences. We saw that participants greatly differed in their response to the score, potentially owing to different reward sensitivities \cite{leue2008reinforcement} or aesthetic preferences \cite{nath2024complexity}. This leads to a diverse area of study that demands larger sample sizes and participant-specific interventions. Our work examined design decision making under reward in the general population rather than in design professionals; effects may be mediated by expertise, which can change intrinsic criteria for judging goal progress or willingness to engage with feedback \cite{waldron1996expertise}. Future work can use reinforcement learning-based computational modeling approaches for phenotyping and personalised support.
\end{itemize}

\section{Conclusion}

Using a computational MDP framework, we formalised design as a measurable sequential decision process and studied the impact of precise, stepwise rewards on design behaviour and perceived experience. Reward maximisation behaviour depended on feedback quality (goal-aligned vs goal-agnostic), and perceived usefulness depended on the nature of the goal (\emph{cheerful}/\emph{dependable} vs \emph{unique}). These findings provide mechanistic insight into how rewards can be used to steer design decision making and offer guidance in how reward feedback should be designed to more effectively support design creativity, agency, and wellbeing.

\bibliographystyle{plain}
\bibliography{references}

@article{simon1973structure,
  author={Simon, H. A.},
  title={The structure of ill structured problems},
  journal={Artificial Intelligence},
  volume={4},
  pages={181--201},
  year={1973}
}

@article{dorst2001creativity,
  author={Dorst, K. and Cross, N.},
  title={Creativity in the design process: co-evolution of problem--solution},
  journal={Design Studies},
  volume={22},
  pages={425--437},
  year={2001}
}

@article{schon1986reflective,
  author={Schon, D. A. and DeSanctis, V.},
  title={The reflective practitioner: How professionals think in action},
  journal={Journal of Continuing Higher Education},
  volume={34},
  pages={29--30},
  year={1986}
}

@misc{cooper2010stagegate,
  author={Cooper, R. G.},
  title={The stage-gate idea to launch system},
  year={2010}
}

@misc{valeri2003phase,
  author={Valeri, S. G. and others},
  title={Implementation of the phase review process in new product development: A successful experience},
  year={2003}
}

@inproceedings{fischer1993embedding,
  author={Fischer, G. and others},
  title={Embedding computer-based critics in the contexts of design},
  booktitle={CHI '93},
  year={1993},
  publisher={ACM}
}

@article{son2022creativesearch,
  author={Son, K. and others},
  title={CreativeSearch: Proactive design exploration system},
  journal={Automation in Construction},
  volume={142},
  pages={104502},
  year={2022}
}

@inproceedings{son2022bigexplore,
  author={Son, K. and Kim, K. and Hyun, K. H.},
  title={BIGexplore: Bayesian Information Gain Framework},
  booktitle={CHI Conference},
  year={2022},
  publisher={ACM}
}

@inproceedings{lee2024sketchguided,
  author={Lee, S. W. and others},
  title={Impact of sketch-guided vs prompt-guided 3D generative AIs},
  booktitle={CHI Conference},
  year={2024},
  publisher={ACM}
}

@inproceedings{swearngin2020scout,
  author={Swearngin, A. and others},
  title={Scout: Rapid exploration of interface layout alternatives},
  booktitle={CHI 2020},
  year={2020},
  publisher={ACM}
}

@inproceedings{lee2020guicomp,
  author={Lee, C. and others},
  title={GUIComp: A GUI design assistant with real-time feedback},
  booktitle={CHI 2020},
  year={2020},
  publisher={ACM}
}

@inproceedings{son2024genquery,
  author={Son, K. and others},
  title={GenQuery: Supporting expressive visual search},
  booktitle={CHI Conference},
  year={2024},
  publisher={ACM}
}

@inproceedings{lin2025inkspire,
  author={Lin, D. C.-E. and others},
  title={Inkspire: Supporting design exploration with generative AI},
  booktitle={CHI Conference},
  year={2025},
  publisher={ACM}
}

@inproceedings{wadinambiarachchi2024effects,
  author={Wadinambiarachchi, S. and others},
  title={Effects of generative AI on design fixation},
  booktitle={CHI Conference},
  year={2024},
  publisher={ACM}
}

@inproceedings{anderson2024homogenization,
  author={Anderson, B. R. and Shah, J. H. and Kreminski, M.},
  title={Homogenization effects of large language models},
  booktitle={Creativity and Cognition},
  pages={413--425},
  year={2024},
  publisher={ACM}
}

@article{bernal2015computational,
  author={Bernal, M. and Haymaker, J. R. and Eastman, C.},
  title={Role of computational support for designers},
  journal={Design Studies},
  volume={41},
  pages={163--182},
  year={2015}
}

@inproceedings{nandy2023machine,
  author={Nandy, A. and Goucher-Lambert, K.},
  title={How does machine advice influence design choice?},
  booktitle={Design Computing and Cognition},
  pages={801--818},
  year={2023},
  publisher={Springer}
}

@inproceedings{shireen2011design,
  author={Shireen, N. and others},
  title={Design space exploration in parametric systems},
  booktitle={Creativity and Cognition},
  year={2011},
  publisher={ACM}
}

@article{lee2020creative,
  author={Lee, J. H. and Ostwald, M. J.},
  title={Creative decision-making in parametric design},
  journal={Buildings},
  volume={10},
  pages={242},
  year={2020}
}

@article{cristie2021versioning,
  author={Cristie, V. and Joyce, S. C.},
  title={Versioning for parametric design exploration},
  journal={Automation in Construction},
  volume={129},
  pages={103802},
  year={2021}
}

@misc{lahikainen2024creativity,
  author={Lahikainen, J. and Ady, N. M. and Guckelsberger, C.},
  title={Creativity and MDPs},
  year={2024}
}

@article{ryan2000self,
  author={Ryan, R. M. and Deci, E. L.},
  title={Self-determination theory},
  journal={American Psychologist},
  volume={55},
  pages={68--78},
  year={2000}
}

@article{ryan1982control,
  author={Ryan, R. M.},
  title={Control and information in the intrapersonal sphere},
  journal={Journal of Personality and Social Psychology},
  volume={43},
  pages={450--461},
  year={1982}
}

@book{sutton1998reinforcement,
  author={Sutton, R. S. and Barto, A. G.},
  title={Reinforcement learning: An introduction},
  publisher={MIT Press},
  year={1998}
}

@article{kluger1996feedback,
  author={Kluger, A. N. and DeNisi, A.},
  title={Effects of feedback interventions on performance},
  journal={Psychological Bulletin},
  volume={119},
  pages={254--284},
  year={1996}
}

@article{ivcevic2024ai,
  author={Ivcevic, Z. and Grandinetti, M.},
  title={Artificial intelligence as a tool for creativity},
  journal={Journal of Creativity},
  volume={34},
  pages={100079},
  year={2024}
}

@inproceedings{nandy2025semantic,
  author={Nandy, A. and others},
  title={Semantic properties of word prompts shape design outcomes},
  booktitle={Design Computing and Cognition},
  year={2025},
  publisher={Springer}
}

@article{bellman1957mdp,
  author={Bellman, R.},
  title={A Markovian decision process},
  journal={Journal of Mathematics and Mechanics},
  volume={6},
  number={5},
  pages={679--684},
  year={1957}
}

@inproceedings{ng1999policy,
  author={Ng, A. and Harada, D. and Russell, S.},
  title={Policy invariance under reward transformations},
  booktitle={ICML},
  pages={278--287},
  year={1999}
}

@book{lee2014bayesian,
  author={Lee, M. D. and Wagenmakers, E. J.},
  title={Bayesian cognitive modeling},
  publisher={Cambridge University Press},
  year={2014}
}

@article{hoffman2014nuts,
  author={Hoffman, M. D. and Gelman, A.},
  title={The No-U-Turn sampler},
  journal={JMLR},
  volume={15},
  pages={1593--1623},
  year={2014}
}

@article{cherry2014csi,
  author={Cherry, E. and Latulipe, C.},
  title={Creativity Support Index},
  journal={ACM TOCHI},
  volume={21},
  pages={1--25},
  year={2014}
}

@article{niv2007dopamine,
  author={Niv, Y. and others},
  title={Tonic dopamine},
  journal={Psychopharmacology},
  volume={191},
  pages={507--520},
  year={2007}
}

@book{berlyne1960conflict,
  author={Berlyne, D. E.},
  title={Conflict, arousal, and curiosity},
  year={1960}
}

@article{schon1992designing,
  author={Sch{\"o}n, D. A.},
  title={Designing as reflective conversation},
  journal={Knowledge-Based Systems},
  volume={5},
  pages={3--14},
  year={1992}
}

@article{gero2004situated,
  author={Gero, J. S. and Kannengiesser, U.},
  title={Situated function-behaviour-structure framework},
  journal={Design Studies},
  volume={25},
  pages={373--391},
  year={2004}
}

@article{lebuda2023metacognition,
  author={Lebuda, I. and Benedek, M.},
  title={Creative metacognition framework},
  journal={Physics of Life Reviews},
  volume={46},
  pages={161--181},
  year={2023}
}

@misc{pan2022reward,
  author={Pan, A. and Bhatia, K. and Steinhardt, J.},
  title={Effects of reward misspecification},
  year={2022},
  note={arXiv:2201.03544}
}

@book{cross2021engineering,
  author={Cross, N.},
  title={Engineering design methods},
  year={2021}
}

@article{liu2003concept,
  author={Liu, Y.-C. and Chakrabarti, A. and Bligh, T.},
  title={Ideal approach for concept generation},
  journal={Design Studies},
  volume={24},
  pages={341--355},
  year={2003}
}

@misc{nath2024creative,
  author={Nath, S. S. and Dayan, P. and Stevenson, C.},
  title={Characterising the creative process},
  year={2024},
  note={arXiv:2405.00899}
}

@book{gollwitzer1996goal,
  author={Gollwitzer, P. M. and Moskowitz, G. B.},
  title={Goal effects on action and cognition},
  year={1996}
}

@inproceedings{busch2023decoding,
  author={Busch, S. and Jensen, N. S. and Barros, M.},
  title={Decoding design briefs},
  booktitle={Nordic Design Research Conference},
  year={2023}
}

@article{pugh2016quality,
  author={Pugh, J. K. and Soros, L. B. and Stanley, K. O.},
  title={Quality diversity},
  journal={Frontiers in Robotics and AI},
  volume={3},
  year={2016}
}

@article{leue2008reinforcement,
  author={Leue, A. and Beauducel, A.},
  title={Reinforcement sensitivity theory meta-analysis},
  journal={Personality and Social Psychology Review},
  volume={12},
  year={2008}
}

@article{nath2024complexity,
  author={Nath, S. S. and others},
  title={Relating objective complexity and beauty},
  journal={Psychology of Aesthetics, Creativity, and the Arts},
  year={2024}
}

@inproceedings{waldron1996expertise,
  author={Waldron, M. B. and Waldron, K. J.},
  title={Influence of designer expertise},
  booktitle={Mechanical Design: Theory and Methodology},
  pages={5--20},
  year={1996},
  publisher={Springer}
}

\end{document}